\documentstyle[prl,preprint,aps]{revtex}
\begin{document}
\draft

\title{A Field Effect Transistor based on the Mott Transition in a
Molecular Layer}

\author{C. Zhou, D. M. Newns, J.A. Misewich and P. C. Pattnaik}
\address{IBM T. J. Watson Research Center, P. O. Box 218,
Yorktown Heights, NY 10598}

\date{\today}
\maketitle

\begin{abstract}
Here we propose and analyze the behavior of
a FET--like switching device, the Mott transition field
effect transistor, operating on a novel principle, the Mott
metal--insulator transition.  The device has FET-like characteristics
with a low ``ON'' impedance and high ``OFF'' impedance.
Function of the device is feasible down to nanoscale
dimensions.  Implementation with a class of organic charge transfer
complexes is proposed.
\end{abstract}

\narrowtext
%\begin{multicols}{2}

The Field Effect Transistor -- the key 3--terminal device
in integrated circuit applications -- is
beginning to encounter obstacles to further development.
Organic--channel field effect transistors, which are contemplated
in near--future applications such as smartcards and displays,
have been dogged by poor
performance, directly traceable to the low carrier mobility in organic
semiconductors\cite{garnier}.
The Si FET eventually encounters scaling
limitations\cite{sifet},
as the minimum channel length is estimated to be limited
to 300--400$\AA$.  A very significant question often asked is
whether there exists an alternative to semiconductor--based
FET technology which might offer a solution to such problems.

Materials which might constitute the 2D conducting channel in a FET
are classified according to their 2D sheet conductance
in Fig.~\ref{fig:sheet}.
The conductance of the Si--based FET
channel forms a benchmark,
against which the performance of FET's fabricated with molecular
semiconductor channels compares very unfavorably.
However, there exists a class of materials with comparable conductance
to Si, namely the
poor metals such as the ${\rm Cu O_2}$ plane in cuprate
superconductors\cite{super}, or the organic or synthetic
metals\cite{torrance,soos,iwasa}, which belong to the
class of doped {\it Mott--Hubbard insulators}\cite{mott}.
The unusual intrinsic nature of this class of
materials can be exploited
in controlling their conductivity.
In this paper we evaluate the concept of a switch
based on the novel physical principle
of the Mott--Hubbard metal--insulator
transition.  We shall demonstrate its feasibility
on physical grounds, and propose appropriate materials which could
constitute the channel in the device.

The Mott--Hubbard insulators (unlike conventional band insulators)
have an odd number of electrons per unit cell.	It is insulating due to
a ``Coulomb blockade'' caused by the repulsive Coulomb interaction $U$
between two electrons on the same site.  Assuming one electron per site,
propagation is blocked by the potential barrier $U$ required to
move an electron on to a nearest neighbor site, provided that the
kinetic energy, measured by the intersite transfer integral $t$,
is less than $U$.  Examples of 2D Mott insulators are alkali metal
layers on GaAs\cite{plummer}, the ${\rm CuO_2}$ planes in undoped
cuprate superconductors\cite{super},
and organic charge transfer complexes\cite{torrance,soos}.

The phase diagram for the Mott--Hubbard insulator is well studied
in, for example, cuprate superconductors.  It is possible to
{\it metallize} the insulator by doping with a certain fraction
of carriers (say holes), typically 15$\%$ carriers per site is appropriate.
The holes permit percolation through the lattice without encountering
the barrier $U$.  In addition to cuprates, the synthetic metals
such as TTF--TCNQ are examples of doped Mott-Hubbard insulators,
with 10--40$\%$ of carriers\cite{torrance}.

Using an appropriate Mott insulator material as a conducting channel,
the proposed device operates by gate control of the carrier
concentration in the channel.  If the carriers are holes, the conduction
is termed p--type, if electrons, n--type.
Implementation is via an FET--type device,
termed a Mott Transition Field Effect Transistor\cite{patent} (MTFET), whose
essential active elements are illustrated in Fig.~\ref{fig:fet}.
The channel,
connecting source to drain when the device is ``ON'', and
acting as insulator when ``OFF'', consists of a
monolayer of the Mott--Hubbard insulating material.
The channel is in close proximity to a thin
insulating spacer layer, which separates it from the gate (control)
electrode.

As in a conventional FET, switching occurs by control of the charge in the
channel via the gate voltage\cite{sze}.  In an enhancement--mode (EM) device,
the monolayer is intrinsically insulating; the device default state
is ``OFF''.  For a p--type EM device, application
of a negative gate voltage induces migration of holes from source and
drain into the channel, converting the channel to a p-type
metallic conductor, and thus switching ``ON'' the device.
Conversely, the n--type EM device can be switched ``ON''
by a positive  gate voltage.

An alternative functionality\cite{fun} is
the depletion--mode (DM) device, in which the channel
molecules are intrinsically doped so as to be conducting.
The device default state
is then ``ON''.  If p--type, the device is switched ``OFF'' by application
of a positive gate voltage,
and conversely for n--type.

Let us consider the case
where the active component in the
layer consists of a single type of molecule.
The molecular energy levels
%(illustrated as inset, Fig.\ref{fig:band})
relative to the source--drain Fermi level
are defined as lying at
$- \varepsilon_{l}$, for the $0 \leftrightarrow 1$
transition (ionization level),
and as $- \varepsilon_{l}+U$, for the $1 \leftrightarrow 2$ transition
(affinity level).
Whichever is the lowest lying excitation
will normally define whether the
device is p-- or n-- type.
%if $\varepsilon_{l}$ is lowest, it is p--type (as in the Fig. 2 inset),
%while if $- \varepsilon_{l}+U$ is the lowest--lying state,
%it is n--type.
In the following, we select the p--type enhancement mode case
(as shown in the inset, Fig. \ref{fig:fet}) for analysis in further detail.

In the insulating ``OFF'' state, the device must be able
to sustain a significant source--drain voltage $V_{DS}$ without excessive
tunneling current or dielectric breakdown.  Fig.~\ref{fig:pot}
shows the potential energy
of an electron as a function of distance along the channel, when the
p--type EM device is in the ``OFF'' state, with a potential $-V_{DS}$
applied between the
drain and the source--gate (assumed at same potential).  For most of the channel length,
apart from a ``healing length'' (typically of the magnitude
of one or two molecule	diameters)
 near the drain end, the potential is that of
the source--gate.  It is clear from Fig.~\ref{fig:pot},
that provided the drain potential
lies below the upper Hubbard band,
{\it i.e.}
\begin{equation}
eV_{DS} < U - \varepsilon_l ,
\end{equation}
then the channel should remain insulating.

One can now physically dope the channel by applying a negative
voltage $-V_G$ to the gate electrode relative to
the source and drain electrodes.
The molecular array will then tend to switch to the metallic state
once its lower Hubbard band moves up to
the Fermi level of the leads.
The actual density of available charge carriers in the array
is determined by the
electrostatic equilibrium established at a given gate voltage.
Neglecting electronic kinetic energy, which is a relatively small
effect, the relationship between doping, or the average hole density
per molecule $\delta$,
and gate voltage $V_G$ is:
\begin{equation}\label{eq:doping}
eV_{G} - \varepsilon_l = \delta e^2/C_{mol} ,
\end{equation}
 $C_{mol}$ in Eq.(\ref{eq:doping}) is the
capacitance per molecule
%in the single gate configuration
%shown in Fig.~\ref{fig:fet}. It is determined
%by the Coulomb interactions in the system
and is given by
\begin{equation}\label{eq:Cmol}
C_{mol}^{-1}= \frac{1}{\epsilon}\sum_{i\neq 0}
\left\{
\frac{1}{|{\bf r}_0-{\bf r}_i|}
- \frac{1}{|{\bf r}_0-{\bf r}_i +2{\bf i}( \tilde{d} + R_{mol} )|}
\right\},
\end{equation}
where ${\bf r}_i$ are the molecular centers,
$\tilde{d} \approx \epsilon d_{ins}/ \epsilon_{ins}$, the
remaining notations being indicated in Fig.\ref{fig:fet}.
%In (\ref{eq:Cmol}),

From Eq.(\ref{eq:doping}), it is clear that
$V_T=\varepsilon_l/e$ defines the minimum gate voltage required
to have a nonzero concentration of charge carriers in the molecular
layer. If $\delta=0.15$ is taken as the typical fraction of carriers
required to adequately be in the ``ON'' (metallic) state (based on
cuprate data\cite{super}), the	required ``ON'' gate voltage
can be specified for a given set of parameters.
In Table I we collect some values calculated from Eq's
(\ref{eq:doping})
and (\ref{eq:Cmol}) assuming $\varepsilon_l=0.25eV$.
For practical
application in a CMOS logic environment
the gate voltage must lie below the breakdown
threshold that both the insulating spacer and the molecular channel
can sustain, but lie above about $0.7 eV.$ for $300 K$ operation, fixing
the gate voltage at $\sim 1 V$.
This implies that $U\sim 1 V$, limiting molecular size to something like
$a_{mol} < 12 \AA$, allowing for some screening.
Theoretical estimates suggest that (see Table I),
in order to have the ``ON'' gate voltage limited to around
$\sim 1 V$, molecules in the channel must be sufficiently
large\cite{synthe} ($a_{mol}>6\AA$).  Hence these considerations
limit the molecular size to a range $6 \AA < a_{mol} < 12 \AA$.
Also the dielectric constant of the insulating spacer
between gate and channel must be significantly larger
than that of commonly used silica ($\epsilon_{ins}=4$).
With these considerations in mind,
the best candidates for the conducting channel
of the MTFET device can only be
found in {\it synthetic materials} or
{\it artificially made structures}.

The molecules need to be reasonably
closely packed so that electronic
wave functions between neighboring sites are well overlapped.
Suitable proposed materials are the charge transfer
salts\cite{torrance,soos} $D^+A^-$,
where at least one of $D$ or $A$ is organic;
examples\cite{ktcnq,C60}
are ${\rm K^+ TCNQ^-}$, and ${\rm K^+ C_{60}^-}$.
It appears that these materials,
with their molecule size of $6\sim 12 \AA$,
and having typically an electronic
bandwidth $\sim 0.5eV$ and a screened Coulomb
interaction $U\sim 1eV$, fulfill
the necessary criteria to be used as
the conducting channel in the proposed MTFET.
Taking as example a system with $a_{mol}=10\AA$,
if implemented as the channel in a single gate configuration with a high
dielectric constant
insulating spacer of thickness
$d_{ins}=50\AA$, the data in Table I gives as
the required ``ON'' gate voltage $V_G=0.64V$, and electric
field in the spacer $E=0.89MV/cm$. The latter is well below
the typical breakdown threshold.
On the other hand, if implemented in a dual gate
configuration\cite{patent},
a regular organic insulator such as polyimide or even silica
will perform satisfactorily(Table I), thus the requirement for high  dielectric
constant materials is not essential.
Finally, in applications other than in CMOS environments
where no low operating voltage is required, the data
in first and third rows of Table I shows that it is even
possible to directly use cuprate as the transistor channel,
implemented together with a high dielectric insulating spacer.

Having established switching between the metallic  and insulating
states of the correlated molecular array,
the analysis is straightforward for the
source--drain current in the ``ON'' state. Within the
gradual--channel approximation\cite{sze}, it is given by
\begin{equation}
I_{DS}=\mu_h\left(\frac{W}{L}\right) n
C_{mol} (V_G- \varepsilon_l /e -\frac{1}{2}V_{DS})V_{DS}\: ; \:
\label{eq:Ids}
\end{equation}
in the low biased region where $V_{DS} < V_{sat}$
and $V_{sat}= V_G-\varepsilon_l/e$. Eq.(\ref{eq:Ids}) is derived
assuming that the doped hole mobility $\mu_h$ is constant,
as is justified by the doping-dependence of the conductance
in cuprates\cite{ito}.
In eq.(\ref{eq:Ids}), $W$ is channel width,
$L$ channel length from source to drain,
and $n$ is surface concentration
of molecules.
The $I_{DS}$ saturates when
$V_{DS} > V_{sat}$ and the saturated current
follows:
\begin{equation}\label{eq:Isat}
I_{DS}=\frac{\mu_h}{2}\left(\frac{W}{L}\right) n C_{mol}
(V_{G} - \varepsilon_l /e)^2 .
\end{equation}
The hole concentration
in the saturation regime has dropped to zero
at the drain end of the channel (``pinch effect''),
which occurs first at
$V_{DS} = V_{sat} = V_{G} - \varepsilon_l /e$.

The results (\ref{eq:Ids}--\ref{eq:Isat})
are plotted in Fig.~\ref{fig:IV}
as the I-V characteristics of the
MTFET device.
They resemble conventional FET characteristics.
The trans--conductance per square in the saturation regime
can be written as:
\begin{equation}
G_{sat}= \mu_hne\delta,
\label{eq:Gsat}
\end{equation}
which can be shown
to be on the order of several quanta of conductance $e^2/h$
(or a sheet resistance of a few $K\Omega$),
%in a poor metal,
consistent with the data from  cuprates(see Fig. 1).

Although charge propagates coherently within the channel in the ``ON''
state, we expect that
tunneling occurs at the channel--lead contacts.
Detailed analysis shows however that the (insulating) edge region
at the contact is confined only within one or two molecular
diameters in the ``ON'' state.
Tunneling conductance has been
estimated to be of the same order of magnitude as eq.(\ref{eq:Gsat}),
indicating that the edge effect does not significantly
affect the characteristic low ``ON'' impedance.

There are two possible fabrication approaches to the device.  Simple
deposition of a thick layer of channel material on top of the prepared
gate--insulator/source--drain structure
a) is confined to EM, where the material is insulating and
will not short out the device, and
b) requires that the molecules arrange themselves
satisfactorily at the oxide interface.
A more designer approach is to exploit the rapidly developing self-assembly
technology, enabling formation of a molecular monolayer, which permits
either EM of DM operation.

Finally, arrays of quantum dots can be viewed
as an extreme limit of the artificially made molecular array.
An important distinction should be emphasized however.
In quantum dot arrays seen currently in the literature\cite{dots},
charges propagate via inter--dot tunneling throughout the channel,
instead of only at the channel--lead contacts envisaged in our
device structure, the entire process of electronic transport
is therefore incoherent, which results inevitably in a
high ``ON'' impedance, an unfavorable situation for practical
applications.
Besides, since the artificial molecules or quantum dots
built with current technologies are too large in size (typically
$\sim 10nm.$) to sustain substantial Coulomb interactions,
the Coulomb blockade phenomenon or switching can at most
be operated at extremely low temperatures.

In conclusion, we have analyzed a MTFET device based on a molecular layer
operating
on a novel principle, the Mott metal--insulator transition.
The channel consists of an  array of $\sim 1nm.$ molecules.
The large electronic Coulomb energy of
order $\sim 1 eV$ allows operation at room temperature and at a voltage of
order $0.5\sim 1 eV$. ``ON'' transport is coherent (all molecules are
exactly the same) allowing low 'ON' resistance. High carrier
density permits device function down to $\sim 50$ molecules, i.e.
to device densities of
order $20$ times maximum Si FET density.
These characteristics
 place the proposed device in the category of
a possible solution to the fundamental problems which the computer
switches is expected to encounter in the next 10--20 years.

%\acknowledements
%The authors thanks discussions with every God. This work was supported
%by God.

\begin{table}[htbp]
\begin{center}
\begin{tabular}{||c|c|c|c|c|c|c|c||}\hline
 &$d_{ins}(\AA)$ & $a_{mol}(\AA)$ & $\epsilon$ & $\epsilon_{ins}$
&$\delta$&$E (MV/cm)$&$V_G (V)$\\ \hline
Single gate&50 &   4   &  4  & 40   &	  0.15	     & 4.73 & 2.71 \\ \hline
Single gate&50 &  10   &  4  & 40   &	  0.15	     & 0.89 & 0.64 \\ \hline
Single gate&1000 & 4   &  4  & 400  &	  0.15	     & 0.45 & 4.77 \\ \hline
Single gate&1000 & 10  &  4  & 400  &	  0.15	     & 0.08 & 1.07 \\ \hline
Single gate&100  & 12  &  4  & 4    &	  0.15	     & 4.36 & 4.87 \\ \hline
Dual gate  &20	 & 10  &  4  & 4    &	  0.15	     & 1.30 & 0.90 \\ \hline
\end{tabular}
\end{center}
\caption{The gate voltage required for the ``ON'' state for
various parameter sets in MTFET for both single gate and  dual gate
configurations.}
\label{tab:gate}
\end{table}

\newpage
\begin{figure}[htbp]
\caption{Schematic sheet conductance for
various categories of materials.}
\label{fig:sheet}
\end{figure}
\begin{figure}[htbp]
\caption{Schematic side view of
three--terminal molecular transistor.
Shown in the inset are energy levels
for electrons of individual molecules in the channel}
\label{fig:fet}
\end{figure}
\begin{figure}[htbp]
\caption{
Energy variation along channel in the presence
of a drain--source bias $V_{DS}$ for the p--type device,
calculated treating electrodes and
channel as thin plates, with gate electrode infinite.  Energy levels in eV.
are  $ U=1.5, \varepsilon_l=0.3, V_{DS}=-0.9$, plate separation
$d=2 nm.$, $\epsilon_{ins}=\epsilon=1$.}
\label{fig:pot}
\end{figure}
\begin{figure}[htbp]
\caption {Current $I_{DS}$ versus drain--source voltage
$V_{DS}(=-V)$ in the channel under various gate voltages,
corresponding to source--end doping
$\delta=0.1-0.5$ (from bottom to top curves).
The current $I_{DS}$ and the voltage $V_{DS}$ in the curve are
expressed in terms of $I_0=G^0_{sat}V_0$
and $V_0=e/C_{mol}$ respectively,
where $G^0_{sat}=(W/L)\mu_hne$.
The gate voltage is negative
in the EM device, with its value {\it increases}
as one moves from bottom to top curves; while it is positive in
depletion mode and {\it decreases} from bottom to top curves.
}
\label{fig:IV}
\end{figure}


\begin{thebibliography}{10}
\bibitem{fun}
Device fabrication options follow from functionality; in the EM case,
the monolayer can
be just the terminating layer of a macroscopic crystalline or amorphous
``blob'' of $D^+A^-$ material,
whereas DM functionality demands that the channel
be a true monolayer, with an insulating protective coat.
\bibitem{synthe}
For this very reason, most inorganic
Mott insulators found in nature are not suitable
to be implemented in the MTFET for applications
in a CMOS logic environment.  Their lattice constant
$a_{mol}$ are too small,
which would inevitably lead to a high operating voltage,
if not beyond the breakdown threshold.
\bibitem{C60}
If the kinetic energy, of which a measure is
the intersite hopping integral $t$ in the Hubbard model,
were large enough relative to U, as happens in K$_3$C$_{60}$,
the system can become metallic even for integral occupation,
However the $1:1$ compound KC$_{60}$ is a Class I insulator.
\end{thebibliography}
\end{document}